# Harnessing non-local orbital-to-spin conversion of interfacial orbital currents for efficient spin-orbit torques

*Shilei Ding* [1,2,3], *Andrew Ross* [2,3], *Dongwook Go*[2,4], *Zengyao Ren* [2,3,5], *Frank Freimuth*[2,4], *Sven Becker* [2], *Fabian Kammerbauer* [2], *Jinbo Yang* [1,6,7], *Gerhard Jakob* [2,3], *Yuriy Mokrousov*[2,4], *Mathias Kläui* [2,3,8]*

[1]State Key Laboratory for Mesoscopic Physics, School of Physics, Peking University, Beijing 100871, China

[2]Institute of Physics, Johannes Gutenberg-University Mainz, Staudingerweg 7, 55128 Mainz, Germany

[3]Graduate School of Excellence Materials Science in Mainz, 55128 Mainz, Germany

[4]Peter Grunberg Institut and Institute for Advanced Simulation, Forschungszentrum Jülich and JARA, 52425 Jülich, Germany

[5]School of Materials Science and Engineering, University of Science and Technology Beijing, Beijing 100083, China

[6]Collaborative Innovation Center of Quantum Matter, Beijing, 100871, P.R. China.

[7]Beijing Key Laboratory for Magnetoelectric Materials and Devices, Beijing 100871, P. R. China.

[8]Center for Quantum Spintronics, Department of Physics, Norwegian University of Science and Technology, NO-7491 Trondheim, Norway

*Correspondence to [klaeui@uni-mainz.de]

**Current induced spin-orbit torques (SOT) allow for the efficient electrical manipulation of magnetism in spintronic devices. Engineering the SOT efficiency is a key goal that is pursued by maximizing the active interfacial spin accumulation or modulating the non-equilibrium spin-density that builds up through the spin Hall and inverse spin galvanic effects. Regardless of the origin, the fundamental requirement for the generation of the current-induced torques is a net spin accumulation. Here, we report the large enhancement of the SOT efficiency in TmIG / Pt by capping with a $CuO_x$ layer. Considering the weak spin-orbit coupling (SOC) of $CuO_x$, these surprising findings are explained as a result of orbital current generated from $CuO_x$ in contact with Pt. This interface-generated orbital current is injected into Pt and converted into a spin current due to the large SOC of Pt. The converted spin current decays across the Pt and exerts a "non-local" torque on TmIG. This additional torque leads to a maximum enhancement of the SOT efficiency of a factor 16 for 1.5 nm of Pt at room temperature, thus opening a path to**

**increase torques while at the same time offering insights into the underlying physics of orbital transport which has been elusive so far.**

Spin-orbit torques (SOTs) are a powerful tool in the arsenal of spintronics aimed at realizing spin-based logic devices and nonvolatile memory [1-5]. As a key part of this spin phenomenon, spin-charge interconversion from spin-orbit coupling effects (SOC) has been intensely studied with mechanics including the spin current generation in heavy metals via the spin Hall effect (SHE) [4,5] and the non-equilibrium spin-density generation at interfaces with inversion symmetry breaking via the inverse spin galvanic effect (ISGE) [4,6,7]. The SOTs originate from the exchange interaction of the non-equilibrium spins and local moments and are of particular importance to enable the electrical manipulation of the magnetization. In efforts to enhance the charge-to-spin conversion rate, various types of materials with large SOC have been investigated, such as heavy metals [8], alloys [9], and topological materials [10]. On the other hand, layers with light elements, whose SOC is negligible, are often assumed to play no significant role in generating SOTs.

Recently it has been shown that oxygen treatment of light elements can have a strong influence, highlighting the important role that oxides can play in exploring the underlying physics of SOTs [11-18]. For example, it was demonstrated that oxidation of metallic ferromagnets in heavy metal (HM) / ferromagnet (FM) heterostructures may change the sign of the SOT [11]. Especially, recent experiments show that oxidation can lead to large torques even for relatively light materials like Cu [17, 18]. In these studies, the Cu / $CuO_x$ interface was reported to lead to the torques with different competing theoretical explanations [11, 16]. One possible scenario is the generation of orbital currents as this does not require SOC [19, 20]. However, Cu / $CuO_x$ interfaces are usually not homogeneous for polycrystalline samples due to the natural oxidation being affected by grain boundaries. And to convert orbital currents to spin currents, which can then be used to generate the torques, one could use an additional heavy metal layer that could then lead to an enhancement of the torques if orbital currents play a role.

In this letter, we aim to elucidate the physical origin of the SOT from $CuO_x$ by studying the SOT in TmIG / Pt / $CuO_x$ heterostructures (see Fig. 1(a)). By varying the Pt thickness, we probe the dependence of the SOT efficiency, and we observe a maximum with a 16-fold increase of the efficiency for TmIG / Pt ($t_{Pt}$) / $CuO_x$ compared to TmIG / Pt samples for a particular thickness of the Pt ($t_{Pt}$). By preventing the oxidation of the Cu layer through additional capping, we rule out the possibility for the enhancement originating from the metallic Pt / Cu interface. Instead, from the non-monotonic thickness dependence, we can attribute the large enhancement of the SOT efficiency to an additional spin current in the Pt originating in the non-local orbital-to-spin conversion of the orbital current generated at the Pt / $CuO_x$ interface, where the mechanism is schematically illustrated in Fig. 1(a). The inversion symmetry breaking at the $CuO_x$ interface gives rise to this orbital current through the orbital Rashba-Edelstein effect (OREE), highlighting the crucial role of non-local generation of SOTs for our spintronic devices.

We use perpendicular magnetized thulium iron garnet ($Tm_3Fe_5O_{12}$; TmIG) films that are deposited on (111)-oriented gadolinium gallium garnet ($Gd_3Ga_5O_{12}$, GGG) substrates using pulsed laser deposition (the deposition conditions can be found elsewhere [21]). Two series of samples are studied: Series A Sub. / TmIG (6.5) / Pt ($t_{Pt}$) and Series B with Sub. / TmIG (6.5) / Pt ($t_{Pt}$) / $CuO_x$ (3) (units in nanometer) (see Fig.1(a)). The Pt and Cu layers were deposited by DC sputtering without breaking the vacuum. The Pt thickness $t_{Pt}$ is varied from 0.5 to 7 nm set by the sputtering time and calibrated via X-ray reflectometry. The Cu layer is naturally oxidized in the atmosphere upon removal from the vacuum. The TmIG layers exhibit perpendicular magnetic anisotropy (PMA) with a coercivity of ~1 mT. The saturation magnetization ($M_s$) is $100 \pm 10 \, kA/m$ [22] obtained from SQUID magnetometry measurements and the in-plane saturation field is ~45 mT [23]. The films were subsequently patterned and etched into 6 μm × 30 μm Hall bars by photolithography and ion milling (inset of Fig. 1(b)). All measurements were carried out at room temperature.

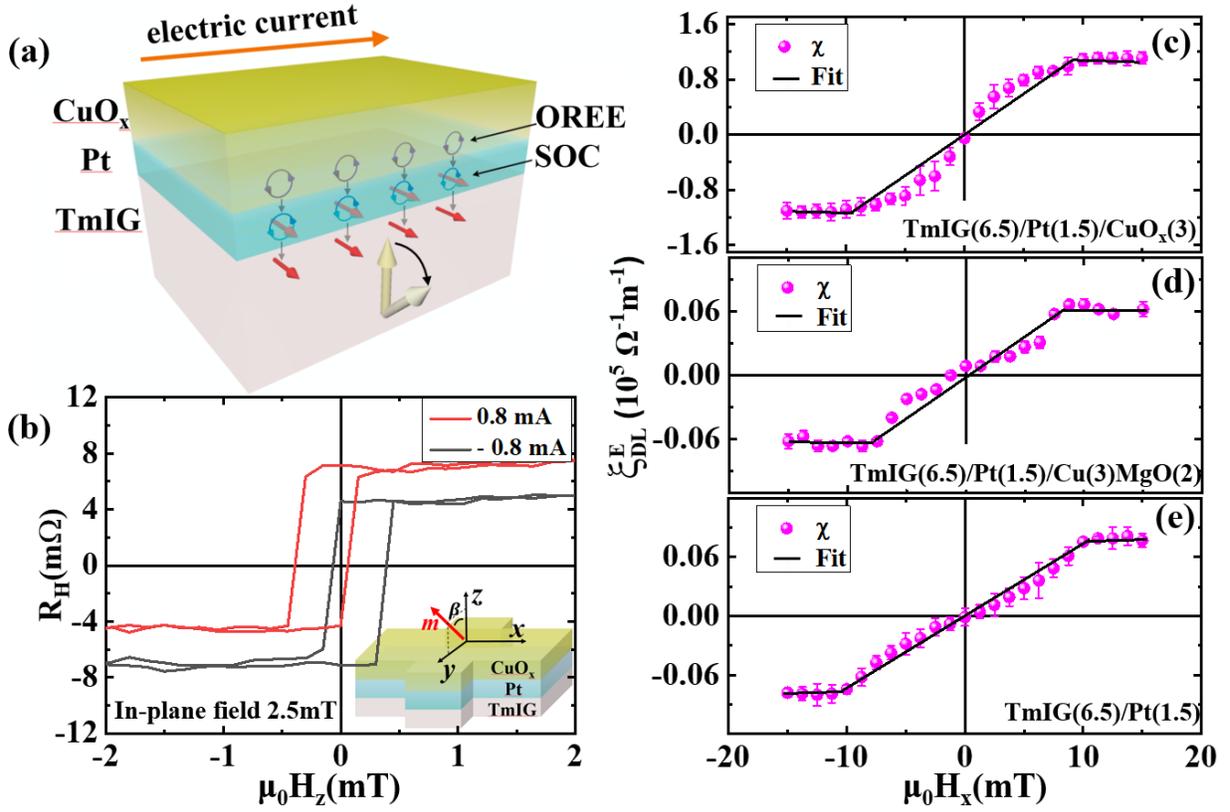

Figure 1. (a) Schematic illustration of non-local generation of SOTs in TmIG / Pt / CuO$_x$ structures. The orbital angular momentum (indicated by blue circulations) is generated via the OREE at Pt / CuO$_x$ interface and injected into the Pt. By the large SOC of the Pt, the orbital angular momentum is converted to the spin (indicated by red arrows), which diffuses across the Pt and exerts a torque on the local moments (yellow arrow) of the magnetic TmIG layer. (b) The shift of the anomalous Hall hysteresis with an in-plane field of $\mu_0 H_x = 2.5$ mT for TmIG (6.5) / Pt (1.5) / CuO$_x$ (3). The probing current is 0.8 mA with different polarities, vertically shifted for clarity (the applied electric field $E = 2.02 \times 10^4$ V/m), the inset shows the scheme of multilayer structure and Hall bar. (c)-(e) The dependence of the effective SOT efficiency $\xi_{DL}^E$ on $\mu_0 H_x$ for TmIG (6.5) / Pt (1.5) / CuO$_x$ (3), TmIG (6.5) / Pt (1.5) / Cu (3) / MgO (2) and TmIG (6.5) / Pt (1.5) respectively. $\xi_{DL}^E$ is calculated from the horizontal displacement in (b) for different probing electric fields at different in-plane fields.

There are two torques present acting on the magnetization of the TmIG, the damping-like torque $\tau_{DL}$ is proportional to $m \times (m \times \sigma)$, and the field-like torque is proportional to $m \times \sigma$ where $\sigma$ is the polarization of the interfacial spin accumulation, and $m$ is the magnetization vector of the magnetic film. The damping-like torque is more relevant for the control of magnetic switching, where it induces an effective out-of-plane field inside Néel type domain walls (DWs), akin to the exchange bias field in PMA exchange bias systems [24]. This torque acts to drive Néel DWs in a direction depending on the DW chirality [25,26]. If there is an interfacial Dzyaloshinskii–Moriya interaction (DMI) present, the DW chirality can be fixed and the DW motion direction depends on the current polarity [25]. By applying strong in-plane fields that reverse the DW chirality, the strength of the DMI can be determined in conjunction with charge current-induced SOTs [26].

To determine the SOTs, we start by measuring the transverse Hall resistance $R_H$ (Fig. 1(b)) as a function of the out-of-plane magnetic field $\mu_0 H_z$. The resistance of the HM is directly related to the relative orientation of $m$, giving rise to hysteresis loops through the anomalous Hall effect as we previously reported for these samples [21]. When $\mu_0 H_x$ is applied in conjunction with $\mu_0 H_z$, this hysteresis loop is shifted by a value $\mu_0 H_L$ to positive or negative fields depending on the sign of the probing current [26]. This shift is shown in Fig. 1(b) for TmIG (6.5) / Pt (1.5) / CuO$_x$ (3) with a probing current $I_C = \pm 0.8$ mA corresponding to an applied electric field $E = V/L = \pm 2.02 \times 10^4$ V/m, where V is the voltage applied along the Hall bar of length L. By studying the value of $\mu_0 H_L$ as a function of $E$, one can obtain the effective spin-torque efficiency per unit applied electric field [21, 26, 27]:

$$\xi_{DL}^E = \frac{2e}{\hbar} M_s t_{FM} \mu_0 H_L / E \qquad (1)$$

Where $M_s$ and $t_{FM}$ are the saturation magnetization and thickness of TmIG film. The value of $\mu_0 H_x$ at the saturation point is called the effective DMI field ($\mu_0 H_{DMI}$) which can be used to extract the value of the DMI constant of the system as previously used for similar samples [21].

We can then determine the value of $\xi_{DL}^E$ by measuring the shift of the hysteresis loop as a function of the electric field that injects the current for differing values of $\mu_0 H_x$ shown in Figs. 1(c)-1(e). The amplitude of $\xi_{DL}^E$ is expected to be linear for small values of $\mu_0 H_x$ before saturating, from which we estimate $\mu_0 H_{DMI} = 10$

± 1 mT, 9 ± 1 mT and 10 ± 1 mT for TmIG (6.5) / Pt (1.5) / CuO$_x$ (3), TmIG (6.5) / Pt (1.5) / Cu (3) / MgO (2) and TmIG (6.5) / Pt (1.5) respectively.

While the DMI field is similar in all samples, the addition of a CuO$_x$ layer leads to a 16 fold increase in the maximum SOT efficiency from $\xi_{DL}^E = 0.07 \times 10^5 \, \Omega^{-1} \text{m}^{-1}$ to $\xi_{DL}^E = 1.16 \times 10^5 \, \Omega^{-1} \text{m}^{-1}$. For the TmIG / Pt sample, the SOTs generated by the SHE induced spin accumulation have been previously reported [28-31]. However, we find here a significant increase in the efficiency and the origin of this large enhancement of $\xi_{DL}^E$ by the CuO$_x$ capping needs to be understood. To check the origin, we firstly add a MgO capping layer to prevent the oxidation of the Cu. In this case, we do not observe an enhancement confirming it is indeed the CuO$_x$ that leads to the enhancement of the SOT efficiency (Figs. 1(c)-1(d)). We also find for our chosen Cu thickness of 3 nm that the entire Cu layer is oxidized, as demonstrated by the large resistance of a TmIG (6.5) / CuO$_x$ (3) Hall bar, thus highlighting it is the Pt / CuO$_x$ interface that is important for the observed enhancement.

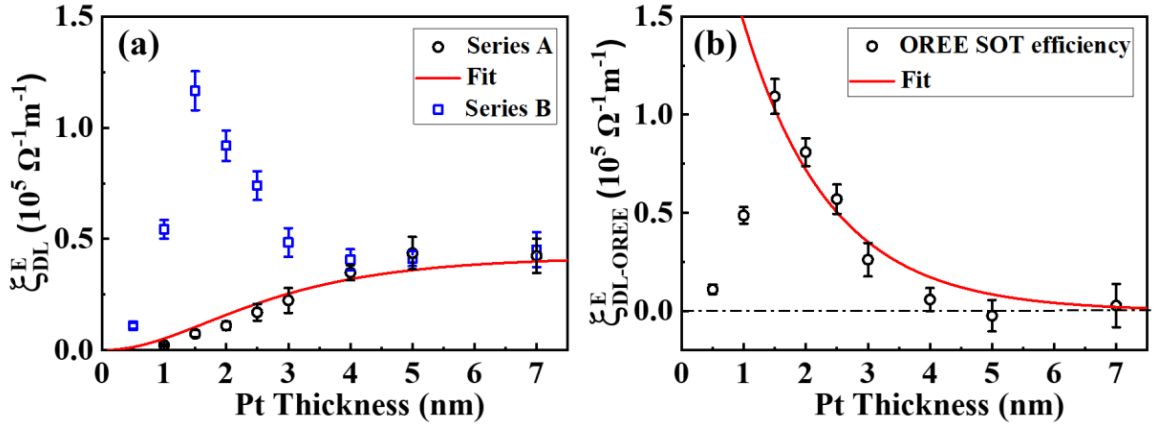

Figure 2. (a) Effective SOT efficiency as a function of Pt thickness for sample Series A without CuO$_x$ (black, circle) and sample Series B with CuO$_x$ (blue, square). (b) The OREE contribution to the SOT efficiency in sample Series B as a function of $t_{Pt}$ ($\xi_{DL}^E = \xi_{DL-SHE}^E + \xi_{DL-OREE}^E$).

Next, we systematically study the evolution of $\xi_{DL}^E$ with varying Pt thickness as shown in Fig. 2(a). For Series A without CuO$_x$, $\xi_{DL}^E$ originates from the SHE, which can be identified as a SHE-induced damping-like-SOT ($\xi_{DL-SHE}^E$). $\xi_{DL-SHE}^E$ increases monotonically with $t_{Pt}$, saturating at $t_{Pt} > 4$ nm due to the finite spin diffusion length in the Pt. If we consider an ideal interface with no spin backflow, this data can typically be fitted by a simple functional form [32]:

$$\xi_{DL-SHE}^{E} = \frac{2e}{\hbar}\sigma_{SH}\left(1 - \text{sech}(t_{Pt}/\lambda_{sf})\right) \quad (2)$$

where $\sigma_{SH}$ is the spin Hall conductivity and $\lambda_{sf}$ is the spin diffusion length in the Pt. A fitting curve corresponding to Eq. (2) is shown in Fig. 2(a), from which we obtain an effective spin diffusion length of $\lambda_{sf} = 1.8 \pm 0.3$ nm, and $\sigma_{SH} = (4.2 \pm 0.3) \times 10^4 \, [\hbar/2e]\Omega^{-1}\text{m}^{-1}$. The effective spin Hall angle can be estimated to be $\theta_{SH} = (2e/\hbar)\sigma_{SH}\rho_{Pt} = 0.010 \pm 0.001$ for a resistivity $\rho_{Pt} = 2.4 \times 10^{-7}$ Ω·m. This value of $\theta_{SH}$ is consistent with previous reports [21, 28]

The Series B samples with CuO$_x$ capping layer demonstrate a completely different dependence on $t_{Pt}$; $\xi_{DL}^{E}$ first increases and then decreases, regaining the same value at $t_{Pt} > 4$ nm as Series A without the CuO$_x$. As the resistivity of the naturally oxidized Cu is orders of magnitude larger than that of Cu [12], the majority of the applied current flows through the Pt itself. Considering the weak SOC and broken inversion symmetry of CuO$_x$ [17], the marked change in the net SOT could come from a combination of the spin current generated from the bulk SHE (Pt) and from an interfacial effect between Pt and CuO$_x$. Meanwhile, Series A samples have only the contribution from the spin current generated via SHE in the Pt.

Our experiments highlight the crucial role of the Pt / CuO$_x$ interface, which is also sensitive to the interfacial cross-sectional area. Although the Pt / CuO$_x$ interface is not directly adjacent to the FM interface, being separated by the Pt layer, it still interacts with the FM and can exert a so-called "non-local" SOT, which underlines the fact that currents generated at a remote non-magnetic interface can exert torques on adjacent ferromagnets [33]. Recently, the effect of the non-local SOT has been interpreted in terms of the orbital hybridization, with a corresponding OREE at the interfaces with Cu inferred to be responsible for the large SOT observed in FM / Cu / CuO$_x$ [17] and FM / Cu / AlO$_x$ [18] trilayers. According to this scenario, orbital angular momentum is generated via the OREE at the Cu / oxide interface. This is analogous to an interface-generated spin current [34], but without hinging on the effect of spin-orbit interaction. The interface-generated "orbital" current is then injected into the FM, where the orbital angular momentum needs to be converted into a spin current by SOC. However, in our experiment, Pt, with its large spin-orbit interaction, automatically converts the generated orbital current into a spin current, such that a spin current rather than the orbital current reaches the TmIG layer and then exerts an effective torque there (Fig. 1(a)). In our case, the electronic structure of Pt, which is crucial for the microscopics of orbital-to-spin conversion [35, 36], is optimal for the purpose, as it features an almost one-to-one ratio of the orbital to spin Hall conductivities [20]. Furthermore, the suppression of the SOT efficiency in Series A for a small thickness of the Pt (Fig. 2(a)) can be attributed to the reflection of the spin current at the bare Pt interface. This reflected spin current then interferes destructively with the primary spin current of interest. However, in Series B, there is no longer a Pt/vacuum interface to reflect the spin current. Instead, the Pt / $CuO_x$ interface leads to a reduction in the reflected spin current due to stronger spin-orbit scattering and additional spin memory loss that arises at Cu based interfaces, and the spin current is efficiently absorbed rather than reflected.

We can then write the net contribution to the SOT efficiency for Series B as $\xi_{DL}^E = \xi_{DL-SHE}^E + \xi_{DL-OREE}^E$. Considering this together with Series A allows us to define the pure orbital-to-spin current conversion contribution $\xi_{DL-OREE}^E$ as a function of Pt thickness. We show this for Series B in Fig. 2(b). The sputtering process normally leads to an island growth mode, and we find that 0.5 nm Pt is not conductive, which indicates the Pt film is not continuous, further reducing the cross-sectional Pt / $CuO_x$ interface area. As $t_{Pt}$ increases up to 1.5 nm, this interfacial area correspondingly increases, and we subsequently observed an increase of $\xi_{DL-OREE}^E$ with $t_{Pt}$. When Pt is thicker ( $t_{Pt}$ >1.5 nm), we expect a saturated value for the orbital-to-spin current conversion in Pt. However, we observe a decrease in the effective $\xi_{DL-OREE}^E$ with $t_{Pt}$ (Fig. 2(b)). We then seek to explain this by considering the decay of the spin current in the bulk Pt. The spin current converted from the orbital current injected into Pt diffuses inside the Pt where it suffers dephasing due to the large SOC of Pt. This diffusion can be estimated as $A exp(-t_{Pt} / \lambda_{sf})$, where a spin current dephasing occurs across the thickness of the Pt from the Pt / $CuO_x$ interface and leads to an additional component to the spin accumulation at the TmIG / Pt interface. We fit the decay in Fig. 2(b) to extract an estimation for $\lambda_{sf}$ of 1.6 ± 0.3 nm. This is, within the error bars, the same value as obtained from Series A samples. Meanwhile $\xi_{DL-OREE}^E$ drops to zero for thicker Pt, highlighting that this enhancement is an interfacial effect. It is worth noting that $\xi_{DL}^E = \xi_{DL-SHE}^E$ of TmIG (6.5) / Pt (0.5) should be smaller than the value from TmIG (6.5) / Pt (1), and we can make a rough estimate that $\xi_{DL}^E = \xi_{DL-SHE}^E + \xi_{DL-OREE}^E \sim \xi_{DL-OREE}^E$ for TmIG (6.5) / Pt (0.5) / CuO$_x$ (3).

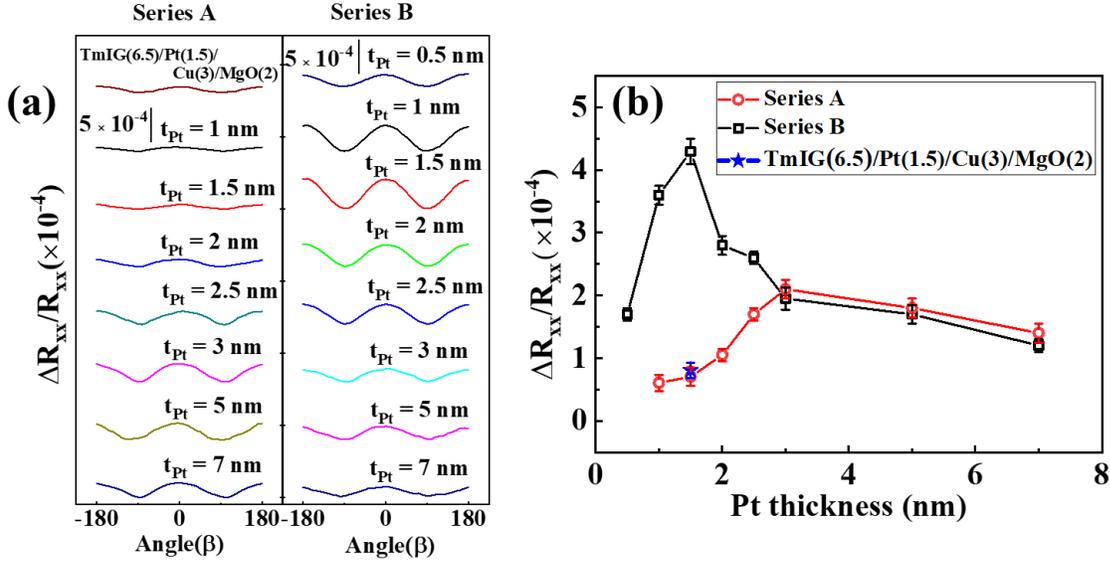

Figure 3. (a) Spin Hall magnetoresistance measurements in the *yz* plane for Series A TmIG (6.5) / Pt ($t_{Pt}$) and Series B TmIG (6.5) / Pt ($t_{Pt}$) / CuO$_x$ (3) samples at an external field of 100 mT. ($\Delta R_{xx}/R_{xx}$ is defined in the main text) (b) The Pt thickness dependence of the SMR ratio $\Delta R_{xx}/R_{xx}$.

To corroborate the enhancement of the SOT for samples with CuO$_x$ capping layers, we next study the effects on the typical spin Hall magnetoresistance (SMR) measurements [37]. The longitudinal resistance ($R_{xx}$) changes with respect to the magnetization orientation of the magnetic layer, related to the *transmission* and/or *reflection* of the spin current. The SMR ratio in *yz* plane is defined as $\Delta R_{xx}/R_{xx} = (R_{xx}(H_z) - R_{xx}(H_y))/R_{xx}(H_y)$, where the $R_{xx}(H_z)$ and $R_{xx}(H_y)$ refer to the longitudinal resistance when the magnetization is saturated along *z* and *y* directions [37]. We show the longitudinal resistance variation $\Delta R_{xx}/R_{xx}$ for the samples with and without CuO$_x$ capping layers in Fig. 3(a). The SMR measurements were performed with the magnetic field rotated in the *yz* plane through an angle β (inset of Fig. 1(b)), in order to exclude the possibility of anisotropic magnetoresistance and isolate a pure SMR contribution [38]. For Series A without CuO$_x$, the Pt thickness dependence of the SMR ratio is consistent with the previous reports on SMR of YIG / Pt [39], where the maximum value of $\Delta R_{xx}/R_{xx}$ occurs at around 3 nm of Pt. When the CuO$_x$ capping layer is introduced, the SMR ratio is enhanced. For sample TmIG (6.5) / Pt (1.5), the SMR ratio is one order of magnitude larger with $CuO_x$ capping. Considering that $\Delta R_{xx}/R_{xx} \propto \theta_{SH}^2$ [38], we would expect the value of $\Delta R_{xx}/R_{xx}$ as a function of $t_{Pt}$ to show the same trend as SOT efficiency, and clearly, both of the spin currents converted from the orbital current and the spin current generated via SHE of Pt contribute to the SMR ratio. Next, to check again whether the observed effect is due to Cu or $CuO_x$, we compared the SMR for TmIG (6.5) / Pt (1.5) / Cu (3) / MgO (2) and TmIG (6.5) / Pt (1.5). The nearly identical values indicate that the Pt / Cu interface is not the source of enhanced SOT and SMR while $CuO_x$ is indeed crucial for the larger SMR signal.

In summary, we observed an up to 16 fold enhancement of the SOT efficiency at room temperature in TmIG / Pt / $CuO_x$ by an additional spin current originating from the conversion of an orbital current, which arises due to the OREE at the remote interface Pt / $CuO_x$. The converted spin current decays across the Pt layer, and it further accumulates at the TmIG / Pt interface, together with the intrinsic SHE of the Pt, thus leading to an increase of the effective spin accumulation compared to the spin current generated by a heavy metal only. The enhanced spin accumulation interacts with the magnetic moment of TmIG via exchange coupling, and we observe an enhancement of the SOT efficiency. The maximum strength of observed SOT efficiency is about 16 times larger at 1.5 nm Pt compared with the sample without $CuO_x$ capping. Our experimental results indicate that the orbital current is generated at the Pt / $CuO_x$ interface and contributes to the large SOT efficiency. Together with utilizing heavy metals like Pt as a conversion layer, our results highlight how orbital angular momentum in oxide-based SOT devices can be harnessed by designing an appropriate stack including a heavy metal layer to convert the orbital current into a spin current. Through optimizing the thickness of the heavy metal layer capped by $CuO_x$, it is possible to significantly enhance the SOT, triggering the potential application of the OREE to low-power spin-orbitronic devices.

Acknowledgments

We acknowledge the support from the Graduate School of Excellence Materials Science in Mainz (MAINZ) DFG 266, the MaHoJeRo (DAAD Spintronics network, Project No.57334897), Deutsche

Forschungsgemeinschaft (DFG, German Research Foundation) – Spin+X TRR 173 – 268565370 (projects A01, A11 and B02), DFG project 358671374, Research council of Norway (QuSpin Center 262633), ERATO "Spin Quantum Rectification Project" (Grant No. JPMJER1402), National Key Research and Development Program of China (Grant No. 2017YFA0206303, 2016YFB0700901, 2017YFA0403701). National Natural Science Foundation of China ( Grant No. 51731001, 11675006, 11805006, 11975035), Max Planck Graduate Center with the Johannes Gutenberg-Universität Mainz (MPGC).